# Itinerant Orbital Hall Effect Mechanism Leading to Large Negative Orbital Torques from Light Metal Vanadium

*Nikhil Vijayan, Durgesh Kumar, Ao Du, Mirco Sastges, Lei Gao, Zijie Xiao, Dongwook Go, José Omar Ledesma-Martin, Hai I. Wang, Daegeun Jo, Peter M. Oppeneer, Rahul Gupta, Gerhard Jakob, Sachin Krishnia*, Yuriy Mokrousov, Mathias Kläui**

Nikhil Vijayan, Durgesh Kumar, Ao Du, Mirco Sastges, Dongwook Go, José Omar Ledesma-Martin, Rahul Gupta, Gerhard Jakob, Sachin Krishnia, Yuriy Mokrousov, Mathias Kläui

Institute of Physics, Johannes Gutenberg University Mainz, Staudingerweg 7, 55128 Mainz, Germany

*Email: skrishni@uni-mainz.de

*Email: klaeui@uni-mainz.de

Mirco Sastges, Dongwook Go, Yuriy Mokrousov

Peter Grünberg Institut and Institute for Advanced Simulation, Forschungszentrum Jülich and JARA, 52425 Jülich, Germany

Lei Gao, Zijie Xiao, Hai I. Wang

Max Planck Institute for Polymer Research, Mainz 55128, Germany

Dongwook Go

Department of Physics, Korea University, Seoul 02841, South Korea

José Omar Ledesma-Martin, Gerhard Jakob, Mathias Kläui

Max Planck Graduate Center Mainz, 55122 Mainz, Germany

Daegeun Jo, Peter M. Oppeneer

Department of Physics and Astronomy, Uppsala University, P. O. Box 516, SE-75120 Uppsala, Sweden

Daegeun Jo, Peter M. Oppeneer

Wallenberg Initiative Materials for Sustainability, Uppsala University, SE-75120 Uppsala, Sweden

Mathias Kläui

Center for Quantum Spintronics, Department of Physics, Norwegian University of Science and Technology, NO-7491 Trondheim, Norway

**Funding:** DFG (Spin+X (A01, A11, B02) TRR 173-268565370 and Project No. 358671374; the European Union's Horizon 2020 research and innovation programme under grant agreement No 863155 (s-Nebula); the European Research Council (ERC) under the European Union's Horizon 2020 research and innovation programme (Grant No. 856538, project "3D MAGiC"); the Research Council of Norway through its Centers of Excellence funding scheme, Project No. 262633 "QuSpin"; the European Union's Horizon Europe programme (EU HORIZON-CL4-2021-DIGITAL-EMERGING-01-14 under grant agreement No. 101070290 (NIMFEIA); and the EU HORIZON-EIC-2023-PATHFINDEROPEN-01-01 under grant agreement No. 101129641 (OBELIX)).

Swedish Research Council (VR) for the International Postdoc VR grant (Grant ID: 2023-06605).

Swedish Research Council (VR), the Knut and Alice Wallenberg Foundation (Grants No. 2022.0079 and 2023.0336), and the Wallenberg Initiative Materials Science for Sustainability (WISE) funded by the Knut and Allice Wallenberg Foundation.

### Keywords

spin-orbitronics, negative orbital torques, intrinsic orbital Hall effect, itinerant circulations

### Abstract

The orbital Hall effect (OHE) has attracted significant attention for developing energy-efficient electronic devices. However, utilizing it in fast, low-power devices requires an enhanced understanding of underlying extrinsic and intrinsic contributions to OHE at timescales ranging from quasi-static to picoseconds. Here, we investigate OHE in light metal vanadium (V) using a combination of selected measurement schemes, spanning the full frequency range. We observe a negative damping-like torque efficiency from V, opposite to conventional theoretical predictions, with a magnitude that depends on the adjacent ferromagnet, a dependence that indicates orbital effects. These results, with consistent torque efficiencies across all frequencies, corroborate a negative and intrinsic OHE in V with a large effective orbital Hall conductivity of –(1.44±0.34) ($\hbar/2e$) ($\times 10^5$ $\Omega^{-1}$m$^{-1}$) and a long orbital diffusion length of (15.0±2.5) nm. To explain the observed OHE, we develop a theoretical model incorporating both local and itinerant circulation contributions to OHE. The model agrees excellently with the experimental results, demonstrating that itinerant contributions are essential for a complete physical understanding of intrinsic OHE. Our consistent experimental and theoretical data highlight the importance of itinerant contributions governing the fundamental understanding of intrinsic OHE and the large effects found open pathways for energy-efficient orbitronic devices.

Nikhil Vijayan and Durgesh Kumar contributed equally to this work

Present Address of Hai I. Wang: Nanophotonics, Debye Institute for Nanomaterials Research, Utrecht University, Princetonplein 1, The Netherlands

Present Address of Rahul Gupta: Department of Physics, University of Gothenburg, 41296, Gothenburg, Sweden

## 1. Introduction

Spin-orbit torques (SOTs) present a promising pathway towards next-generation memory and neuromorphic devices, offering superior energy efficiency and endurance comparable to those of spin-transfer torque-based devices[1–5]. Conventionally, the underlying mechanisms that lead to SOTs are the spin current ($I_S$) generated through the spin Hall effect (SHE)[6–8] and/or interfacial Rashba-Edelstein effect (REE)[8–10]. However, despite their pivotal role in enabling efficient SOTs[7,11], significant drawbacks are the reliance on heavy, rare, expensive, and sometimes toxic materials. Therefore, it is essential to explore alternative mechanisms beyond the SHE and/or REE that do not require such material systems.

Recent theoretical studies reinstate the orbital Hall effect (OHE)[12–15], which, unlike the SHE, does not require materials with a large spin-orbit coupling (SOC). Therefore, it enables a broader range of material choices[16,17]. Moreover, the corresponding orbital Hall conductivities (OHCs) are predicted to be orders of magnitude higher than the spin Hall conductivities (SHCs)[17,18]. The OHE refers to the generation of transverse orbital current ($I_L$) when a charge current ($I_C$) is injected into a non-magnet (NM), which can then diffuse into the adjacent ferromagnet (FM) and excite its magnetization[19]. Since $I_L$ cannot directly interact with the FM's local magnetization, a conversion from $I_L$ to $I_S$ is required. This can be achieved either by using an FM with an appropriate SOC[20,21] or by inserting a converter layer (such as Pt) of high SOC between the NM and the FM[15,22,23]. Building on theoretical predictions, several experimental studies have confirmed the existence of OHE in various materials. For instance, the sign and magnitude of the damping-like torque (DLT) were found to depend on the choice of the FM layer[21,24,25], which is attributed to OHE in the NM and subsequent $I_L$ to $I_S$ conversion in the FM. Furthermore, direct evidence of a large OHE was provided through magneto-optical detection of orbital accumulation at the sample surfaces[26–28]. Beyond quasi-DC and microwave frequency (RF) electrical and optical measurements, THz emission studies of similar materials have also revealed signatures of the reciprocal inverse OHE (IOHE)[29–33].

Despite these promising prospects, many materials systems exhibit discrepancies in experimentally measured and theoretically predicted OHCs [27,34–38]. This can arise from theoretical approximations, such as disregarding the itinerant contributions to orbital motion [17,39], or experimental artifacts inherent to certain measurement schemes, operating at a particular timescale, or both. However, to explore the OHE in practical devices, robust theoretical predictions and experimental measurements are crucial. This is because we require materials exhibiting substantial OHE responses for practical applications whose identification depends critically on reliable theoretical predictions. These have so far predominantly captured some of the intrinsic contributions to the OHE. Therefore, a complete physical understanding of the origins of inconsistent results is needed. For this, we need to investigate the OHE through different measurement schemes, operating at timescales ranging from quasi-static to terahertz (THz): The operational timescale of a measurement technique treats the intrinsic and extrinsic contributions to the OHE differently, often resulting in inconsistent reported torque efficiencies[40,41]. At ultra-fast time scales, the driving frequency can exceed the impurity scattering rate, and therefore, in contrast to intrinsic effects, extrinsic scattering contributions will become strongly frequency dependent. In addition, we need to employ our approach to materials with low resistivity values where intrinsic contributions to the OHE are expected to dominate, in particular at ultra-fast time scales. For instance, an experimental study shows that Pt with resistivities ranging from ~20-70 μΩ cm results in dominant intrinsic contributions to the SHE[42,43]. Earlier theoretical work also suggests that in a moderately dirty regime, the intrinsic OHC is independent of the resistivity of the material[44]. On the theoretical side, we need to devise an improved framework that considers the contributions from local as well as itinerant circulations for reliable theoretical OHCs. Therefore, we investigate OHE in light 3*d* transition metal V, exhibiting a low resistive phase ((85.0±2.2) μΩ cm) and particularly apt for capturing the intrinsic contributions. V is predicted to possess comparable OHCs to 4*d* and 5*d* transition metals[16,17], and negligible SHCs owing to the much smaller SOC, making it easy to disentangle the orbital effects from spin effects.

To facilitate a reliable comparison and understand the OHE origin, we study the nature of OHE in V films across a wide frequency range using quasi-static second harmonic Hall (SHH) voltage, gigahertz (GHz) spin-torque ferromagnetic resonance (ST-FMR), and ultrafast THz emission measurements. We find a robust negative DLT efficiency in V across the full frequency spectrum. Further, to ensure the orbital character of the observed torque, we combine $Fe_{60}Co_{20}B_{20}$ (FCB) and Ni with the V and demonstrate the dependence of DLT on the choice of the FM layer. These observations, along with qualitative consistency in torque efficiencies across the entire frequency range, corroborate a negative OHE of intrinsic origin in our V films. Through a V-thickness dependence study, we quantify the large and negative effective OHC and the long orbital diffusion length. The observation of negative OHE in our V samples is opposite to the results of conventional previous theoretical predictions[17,39]. To understand the underlying contributions to observed negative OHC, we construct a Wannier-based theoretical framework, which not only considers the conventional local contributions but also includes previously neglected itinerant circulation contributions, fundamentally related to the sample surface. Our experimentally estimated and theoretically calculated OHCs agree well, highlighting the important role of itinerant circulations in capturing the full physical picture of OHE[45–47].

## 2. Results and Discussion

First, we quantify the orbital torques in substrate//Ta(3)/NM(2)/Pt(1.5)/FM(2.6)/Pt(1.5), where NM = V or Pt and FM = $[Co(0.2)/Ni(0.6)]_{\times 3}/Co(0.2)$, samples using quasi-static SHH voltage measurement technique[48] (numbers in parentheses are thicknesses in nm) (Methods and supplementary information (SI) S1). The top Pt layer is introduced to compensate for the SHE contribution from the bottom Pt layer. During SHH voltage measurements, an alternating current is injected along the $x$-axis (Figure 1a), and the 1st and 2nd harmonic components of the transverse Hall voltage are simultaneously measured while sweeping an external magnetic field ($B_{ext}$). Figure 1b shows the measured 1st and 2nd harmonic components of the Hall voltages for the V/Pt/FM/Pt sample. We fit the measured 1st harmonic signal using the Stoner–Wohlfarth model[49] and determine the anisotropy field, the angle between $B_{ext}$ and the sample plane, and the orientation of the magnetization at different $B_{ext}$. Subsequently, we fit the measured 2nd harmonic component of the Hall voltage using Equation S5 (for details, see SI S2) to extract the DL effective field ($B_{DL}$)[48,50] at a certain charge current density ($J_C$). To determine the sign of torque efficiency in these samples, we measure the DLT efficiency in Pt(3.5)/FM/Pt(1.5) reference sample and observe an opposite sign (Figure 1c). The $B_{DL}$ scales linearly with $J_C$ for both samples (Figure 1d), and the opposite slopes for Pt and V samples further confirm the negative DLT in V-based sample. From $B_{\mathrm{DL}}$ *vs* $J_{\mathrm{C}}$, the DLT efficiency per unit electric field ($\xi_{\mathrm{DL}}^{E}$) can be estimated using[51]

$$\xi_{\mathrm{DL}}^{E} = \frac{2e}{\hbar} M_s t_{\mathrm{FM}} \frac{B_{DL}}{E} \tag{1}$$

where $e$, $\hbar$, $M_{\mathrm{S}}$ (SI S3 for $M_s$ values), and $t_{\mathrm{FM}}$ represent electronic charge, reduced Planck constant, saturation magnetization, and FM layer thickness, respectively. Besides, $E = \rho_{xx} J_c$, with $\rho_{xx}$ being the longitudinal stack resistivity (SI S4) of the Hall bar structure. By normalizing the electric field, we avoid the complexities of estimating the current shunt effects across the NM layers. The $\xi_{\mathrm{DL}}^{E}$ for samples with NM = V and Pt is found to be –(0.35±0.02) $\times 10^5\ \Omega^{-1}\mathrm{m}^{-1}$ and (1.12±0.05) $\times 10^5\ \Omega^{-1}\mathrm{m}^{-1}$, respectively.

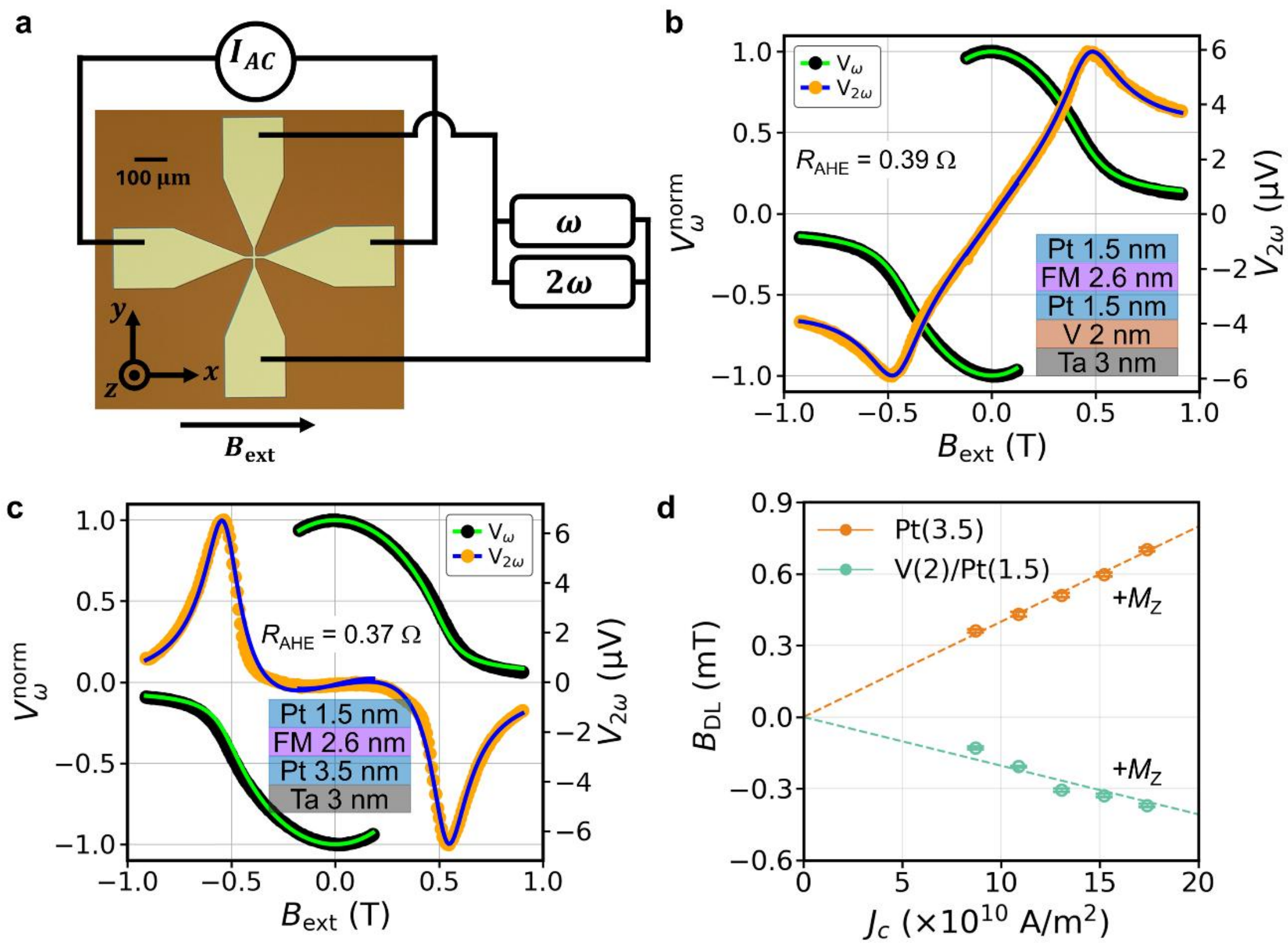

**Figure 1.** Quantification of damping-like torque efficiency in V and Pt-based samples. **a)** Schematic representation of the harmonic Hall voltage measurement setup. 1st harmonic (normalized) and 2nd harmonic Hall voltages (at $J_{\mathrm{C}} = 1.74 \times 10^{11}$ A/m$^2$) and corresponding fits to extract damping-like effective field are shown for V/Pt/FM/Pt **b)** and for Pt/FM/Pt **c)** samples. The insets of figures (**b-c**) represent the corresponding thin film structures. **d)** The amplitude of damping-like effective fields as a function of current density for V/Pt/FM/Pt and Pt/FM/Pt samples, indicating that the sign of DLT efficiency in V is opposite to that due to the SHE in Pt.

The observation of the negative DLT efficiency in V may be attributed to a negative SHE or a negative OHE. A strategy to understand the underlying origin is to investigate the dependence of DLT on the choice of the FM layer. It has been experimentally reported that Ni exhibits more efficient $I_{\mathrm{L}}$ to $I_{\mathrm{S}}$ interconversion due to its large SOC[18,25]. The FCB, however, has a poor $I_{\mathrm{L}}$ to $I_S$ conversion efficiency, predominantly showing signals from the SHE contribution rather than the OHE[25,32]. Therefore, we incorporate Ni and FCB with V and Pt in a Ta(1)/FM(5)/NM(3)/MgO(2)/Ta(2) stack and quantify the DLT efficiencies. We note that in SHH voltage measurements, the FM layer-dependent variation in the planar Hall effect (PHE)[18] may cause errors in the estimation of the torque efficiencies due to magnon contributions[52].

So, to circumvent possible errors caused by the variation in PHE and to corroborate our results, we select a complementary technique to quantify the DLTs in the above-mentioned samples: ST-FMR. This frequently used technique operates at GHz frequencies compared to the SHH. We note that at these frequencies, which are low compared to the charge carrier scattering time, both extrinsic and intrinsic contributions to the OHE are measured, similar to the SHH [53].

In ST-FMR measurements, an RF $I_C$ is injected into the samples. As a result of oscillating torques on the FM due to SHE and/or OHE from the NM and Oersted (Oe) field, the FM generates an oscillating anisotropic magnetoresistance (AMR)[54]. The mixing of the oscillating AMR and $I_C$ produces a rectified DC voltage ($V_{mix}$), which we measure while sweeping an in-plane magnetic field ($B_{ext}$). The measured $V_{mix}$ consists of the symmetric ($S$) and antisymmetric ($A$) Lorentzians, which are related to the $V_{mix}$ through Equation S6 (SI S5)[54,55]. $S$ primarily arises from the DLT generated by the $I_S$ and/or $I_L$ injected into the FM. However, $A$ originates from the torque due to the RF Oe field and the field-like torque (FLT). From the extracted magnitudes of $S$ and $A$ at a given power, frequency, and angle, the ST-FMR efficiency ($\xi_{FMR}$) can be obtained using Equation S7 (SI S5)[55–57]. When the FLT is negligible, the $\xi_{FMR}$ is equivalent to the DLT efficiency per unit current density ($\xi^J_{\mathrm{DL}}$)[25]. The corresponding $\xi^E_{\mathrm{DL}}$ can be estimated using $\xi^E_{\mathrm{DL}} = \xi^J_{\mathrm{DL}}/\rho_{xx}$[25] (see Methods and SI S5).

As shown in Figure 2a-b, the deconvoluted symmetric component exhibits the same sign for FCB/Pt and FCB/V samples. However, the magnitude of $\xi^J_{\mathrm{DL}}$ in FCB/V ($|\xi^J_{\mathrm{DL}}|$ = (0.015±0.001) is approximately six times smaller compared to the same in FCB/Pt ($|\xi^J_{\mathrm{DL}}|$ = (0.093±0.002)). We note that the Ta seed layer may result in non-negligible contributions to overall torque efficiency, and therefore, we measure the Ta(1)/FCB(5)/cap sample for reference and estimate the DLT efficiency to be $|\xi^J_{\mathrm{DL}}|$ = (0.017±0.006) with the same sign as for the FCB/Pt sample (SI S6). Therefore, the DLT due to SHE in V is negligibly small (~0.002±0.006), which is consistent with the theoretical predictions[17].

To find out whether the torque efficiency depends on the FM, which indicates an orbital contribution, we replace FCB with Ni. Interestingly, the sign of the symmetric component in Ni/V is found to be opposite compared to the Ni/Pt sample (Figure 2c-d), with similar magnitudes of torque efficiency ($|\xi^J_{\mathrm{DL}}|$ = (0.020±0.001)). Moreover, we find a negligible contribution from the Ta seed layer ($|\xi^J_{\mathrm{DL}}|$ = (0.006±0.001)) (SI S6). These results confirm the negative sign of DLT in Ni/V. Furthermore, we observe approximately a seven times higher $\xi^J_{\mathrm{DL}}$ in Ni/V samples, as compared to the FCB/V sample, which cannot be attributed to the SHE and therefore confirms the presence of an OHE in V. These results collectively indicate the negative OHE in V. Concerning Ni/Pt, the same sign of $\xi^J_{\mathrm{DL}}$ in FM/Pt is in agreement with the fact that both SHE and OHE exhibit the same sign[21],[17]. The reduction in the magnitude of torque efficiency for Ni/Pt as compared to FCB/Pt may additionally be related to the difference in interface transparency in these samples[58].

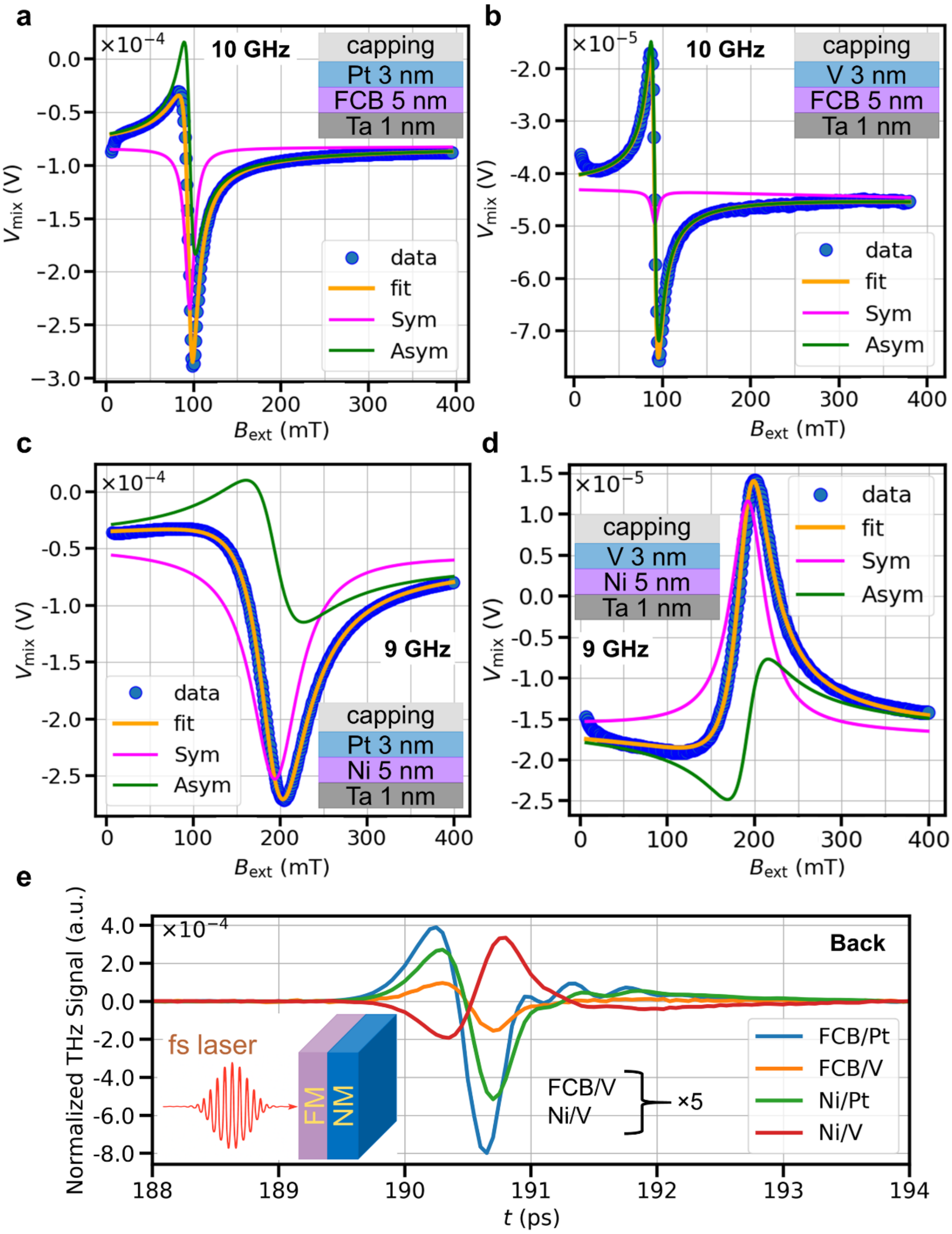

**Figure 2.** Dependence of damping-like torque and THz emission on the ferromagnet for V and Pt-based samples. Experimentally measured ST-FMR data along with the overall fit (orange line) (abbreviated as 'fit'), deconvoluted symmetric (magenta) and asymmetric (green) components for **a)** FCB/Pt, **b)** FCB/V, **c)** Ni/Pt, and **d)** Ni/V samples. Here, 'sym' and 'asym' are the acronyms for symmetric and antisymmetric components of the signal. The insets represent the corresponding film stacks (capping: MgO(2)/Ta(2)). **e)** The THz emission spectra for the identical FM/NM samples when excited from the back (see inset). The data corresponding to FM/V samples have been scaled for better visualization.

It has been experimentally observed that the torque efficiencies extracted using ST-FMR measurements can be influenced by the presence of the spin pumping effect, particularly for stacks with FM of low AMR[59]. This is because the rectified voltage from ST-FMR and the DC voltage from spin pumping are measured within the same electrical circuit. Moreover, the sign and spectrum shape of voltages from spin-pumping and the symmetric component of $V_{mix}$ from ST-FMR are also the same. Similarly, orbital pumping may also influence the measured torque efficiency[60,61].

A way to quantitatively rule out artifacts related to spin/orbital pumping in ST-FMR measurements is to measure THz emission in these samples, and this technique additionally covers the (sub-)picosecond (ps) timescale. In contrast to quasi-static SHH voltage and GHz ST-FMR measurements, the driving frequency may exceed the impurity scattering rate at ultrafast frequencies, and therefore, extrinsic contributions to the OHE will not contribute to the effective measured OHE signal.

In THz emission experiments[62,63], an incident femtosecond (fs) laser pulse on an FM/NM bilayer produces spin and/or orbital current with density $J_S$ and $J_L$ in FM through ultrafast demagnetization. This spin and/or orbital current then diffuses into the NM and gets converted into charge current with density $J_C$, owing to the IOHE/inverse SHE (ISHE) of NM. Consequently, this results in the generation of a THz electric field.

Therefore, we measure the THz emission on the same samples used for the ST-FMR measurements. We perform the measurements both (a) by rotating the direction of $B_{ext}$ and (b) by exciting the samples from the front (film surface) and back (substrate) sides. From the measured THz emission spectra, we focus on the normalized odd-in-magnetization signal to account for THz emission from the spin-orbitronic effects[64] (Methods and SI S7). For the back excitation, the THz emission spectra show the same polarity for FCB/Pt and FCB/V samples (Figure 2e). In contrast, the polarity of the THz emission spectra is reversed in Ni/V as compared to the same for Ni/Pt. These observations are consistent with the ST-FMR results and indicate that the sign of the IOHE in V is opposite to Pt (see SI S8 for the results of front illuminated measurements). Irrespective of the measured sample, the change in relative polarity of the front and back illuminated measurements confirms that the origin of THz emission is of an electric dipole nature (i.e., related to ISHE or IOHE)[63]. Subsequently, we extract the peak-to-peak THz signal strengths and observe that the FCB/Pt and FCB/V samples exhibit the maximum and minimum peak-to-peak THz signal strengths, respectively. Moreover, Ni/V displays a larger peak-to-peak THz signal as compared to the FCB/V sample.

Following the confirmation of a negative OHE in our V samples, it is critical to study the following key aspects for a fundamental understanding of the nature of OHE in V, as well as the practical relevance of our stack presented in Figure 1. These aspects are (*i*) to understand the conversion of $I_L$, generated in the V layer, into $I_S$ in the adjacent Pt layer and (*ii*) to investigate characteristic length scales over which orbital angular momentum (OAM) carriers can propagate. To study the first aspect, we measure the torque efficiency by varying the Pt layer thickness ($t_{Pt}$) in Ta(3)/V(6)/Pt($t_{Pt}$)/FM/Pt($t_{Pt}$) ($t_{Pt}$= 1, 1.5, 2, 3, 4, and 8 nm) samples (Figure 3a). For $t_{Pt} \leq 2$ nm, the torque efficiency remains negative, and the magnitude decreases with the increase in $t_{Pt}$. For $t_{Pt} > 2$ nm, the sign of DLT efficiency reverses to positive, with the magnitude independent of $t_{Pt}$. At smaller thicknesses, the $I_L$ from the V layer is converted into $I_S$ in the Pt layer, which subsequently interacts with the magnetization of the FM layer. As the Pt thickness increases, the converted $I_S$ dephases across the Pt layer, resulting in the reduction of the torque efficiency[22]. With the further increase in $t_{Pt}$, the DLT from the Pt dominates and results in positive torque efficiencies. This is also confirmed through the observation of positive torque efficiency in Ta(3)/V(6)/Pt(4)/FM/Pt(1.5) sample (SI S9). Moreover, the constant torque efficiency at higher $t_{Pt}$ is attributed to the finite spin diffusion length of Pt[22,65]. Importantly, the positive torque efficiency at larger $t_{Pt}$ also confirms that the observed negative torque in our samples can not attributed to the Pt capping layer.

**Table 1.** Comparison of $\xi_{DL}^{E}$ (SHH voltage and ST-FMR measurements) and peak-to-peak THz signal strength. Here, $\lvert\xi_{DL}^{J} - corr.\rvert$ is the $\xi_{DL}^{J}$ after removing the Ta seed layer contributions.

| SHH Voltage Measurements | | | |
|---|---|---|---|
| Sample | | $\lvert\xi_{\mathrm{DL}}^{E}\rvert$ [×$10^4$ $\Omega^{-1}$m$^{-1}$] | |
| Pt(3.5)/FM(2.6)/Pt(1.5) | | 11.2±0.46 | |
| V(2)/Pt(1.5)/FM(2.6)/Pt(1.5) | | 3.5±0.24 | |
| ST-FMR Measurements | | | THz Emission Measurements |
| Sample | $\lvert\xi_{\mathrm{DL}}^{J} - corr.\rvert$ | $\lvert\xi_{\mathrm{DL}}^{E}\rvert$ [×$10^4$ $\Omega^{-1}$m$^{-1}$] | Peak-to-peak THz signal strength [×$10^{-4}$ a.u.] |
| FCB/Pt | (0.076±0.007) | 24.30±1.76 | 11.90±0.37 |
| FCB/V | (0.002±0.006) | 1.71±0.19 | 0.50±0.03 |
| Ni/Pt | (0.025±0.001) | 5.20±0.38 | 7.90±0.08 |
| Ni/V | (0.014±0.001) | 2.35±0.17 | 1.06±0.02 |

To investigate the second aspect, we conduct a V thickness ($t_V$)-dependent study, with the film structure Ta(3)/V($t_V$)/Pt(1.5)/FM/Pt(1.5) ($t_V$ = 2, 3, 6, 12, 24, 30, and 48 nm). The $\xi_{\mathrm{DL}}^{E}$ increases with $t_V$, exceeding the corresponding $\xi_{\mathrm{DL}}^{E}$ for the Pt(3.5)/FM/Pt(1.5) reference sample at higher $t_V$ (Figure 3b). More importantly, the sign of $\xi_{\mathrm{DL}}^{E}$ remains negative for all $t_V$, strengthening our observations of negative DLT efficiency due to OHE in V. Subsequently, we fit the $\xi_{\mathrm{DL}}^{E}$ *vs* $t_V$ data with the following drift-diffusion equation to extract the effective OHC ($\sigma_{OH}^{eff}$) and orbital diffusion length ($\lambda_{OH}$)[26,51],

$$\xi_{DL}^{E} = \sigma_{OH}^{eff}\left(1 - \mathrm{sech}\left(\frac{t_V}{\lambda_{OH}}\right)\right) \tag{2}$$

We extract the $\lambda_{OH}$ for V to be (15.0±2.5) nm, which is significantly larger than the values reported for V[27,38,66]. We find $\sigma_{OH}^{eff}$ is to be –(1.44±0.34) ($\hbar/2e$) (×$10^5$ $\Omega^{-1}$m$^{-1}$).

Our observations through innovative experimental investigations by combining quasi-static and GHz-frequency electrical transport and THz emission measurements corroborate a robust negative OHE in V. Irrespective of the probing time scales and other limitations of the measurement schemes, the magnitudes of $\xi_{DL}^{E}$, obtained from SHH voltage and ST-FMR measurements, and the peak-to-peak THz signal strengths follow the same trend (Table 1). This implies (a) the contribution of different artifacts is qualitatively negligible and (b) most importantly, the OHE in our V samples is of intrinsic origin. The observation of negative intrinsic OHE in our V samples is surprising, given the previous prediction of positive OHE in V from conventional theory[16,17].

To understand the sign of the OHE in V, observed experimentally, we developed a Wannier-based framework, which extends the modern theory of orbital magnetization to the evaluation of the full OAM operator. The developed modern formulation considers contributions to the OAM that goes beyond conventionally used approaches based on atom-centred approximation (ACA), which only considers local atomic effects. The implemented formulation allows us to separate the full OAM operator into contributions due to local electron circulation (LC) and contributions that fundamentally originate in the circulation of Wannier functions at the surface of the sample, referred to as the itinerant circulation (IC). Note that our LC contribution still differs from the ACA even though they both describe local effects, because the LC part of the OAM operator is non-local (i.e., involves matrix elements among Wannier functions centered on different atoms) in our formulation. Excluding all non-local terms from the LC part of the OAM operator yields a similar behaviour to the ACA (SI S10).

So far, owing to the lack of material-specific studies, which take the precise Wannier-based description of the electronic structure into account, it is not fully clear if and how the IC contributions affect the effective OHE in transition metals. So here, we examine the case of bcc vanadium in detail. Our calculations predict that at the Fermi energy of this material, the total LC + IC contribution yields a value which is remarkably close to the experimentally estimated value of $-(1.44\pm0.34)$ $(\hbar/2e)$ $(\times10^{5}\ \Omega^{-1}\text{m}^{-1})$ for $\sigma_{OH}^{eff}$ (Figure 3c and Table 2). Importantly, we observe that the negative sign of the OHC is a result of a competition between LC and IC contributions, which carry an opposite sign. The agreement between experiment and theory is of great interest, as it suggests that the IC contributions to the orbital transport in realistic materials generally cannot be neglected. This calls for future extensive studies of proper OHC in transition metals. Nevertheless, it should be noted that the theoretical method used here only assesses the intrinsic OHC and is still based on approximations like the thermodynamic low-temperature limit. This can still lead to quantitative discrepancies between experiment and theory, especially in materials with large side jump or skew scattering contributions[67] to the OHC. We furthermore acknowledge a limitation of the conventional definition of the OHC, related to the fact that OAM is not a conserved quantity, which can influence physical observables such as orbital accumulation[27] or torque. Altogether, the effective OHC extracted from experiments may deviate substantially from the conventional OHC, and a quantitative OHE identification is challenging.

Given an excellent agreement between our experimental and theoretical results, we discuss possible reasons compared to previous work: A recent study has reported the existence of OHE in V through the Hanle magnetoresistance[38]. In these results, the reported OHC values were found to be smaller compared to the theoretically predicted values. They attributed this discrepancy to the disorder present in V films. However, our estimated resistivity values of V are much smaller ((85.0 ± 2.2) μΩ.cm for V (3 nm)). Moreover, we find a much higher OHC in our V films. A theoretical study also suggests that in a moderately dirty regime, the intrinsic OHC is independent of resistivity ($\rho$), however, it decreases approximately $\propto \rho^{-2}$ in the high resistivity regime[44]. Therefore, our V samples may be in a moderately dirty regime. Moreover, in our V-thickness dependence measurements, the change in $\xi_{DL}^{E}$ is smaller for $t_V \leq 6$ nm, as compared to the same in the thickness range of 6 nm $\leq t_V \leq$ 24 nm. This observation may be related to the contributions from the sample surface, qualitatively in line with our theoretical calculations. We finally note, though, that with the present understanding of the modern theory of OAM, we cannot quantitatively predict how LC and IC contributions scale with the thickness.

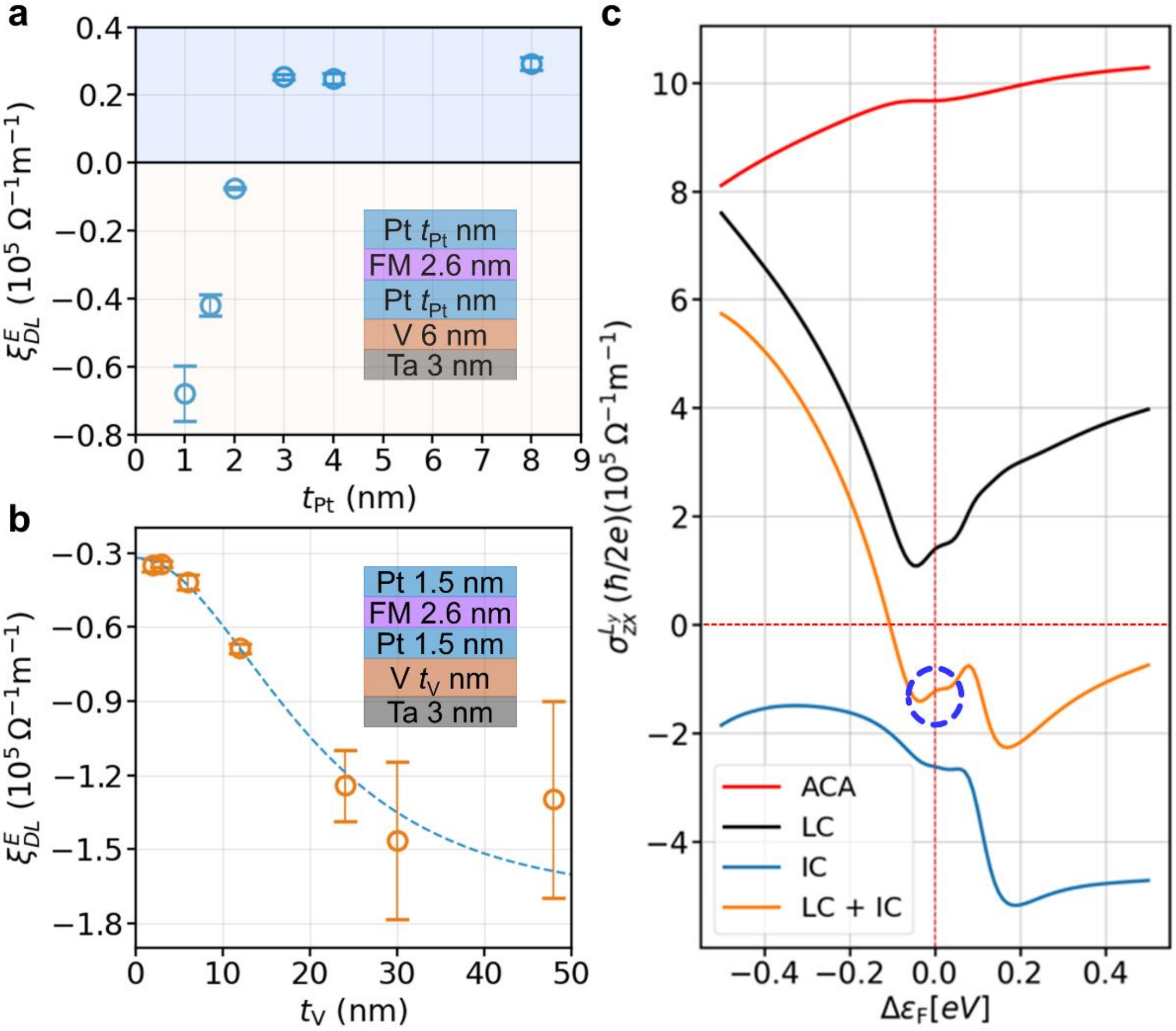

**Figure 3.** Investigation of orbital to spin current conversion, estimation of orbital diffusion length in V, and theoretical ascertainment of negative orbital torque from V. **a)** The $\xi_{DL}^{E}$ as a function of Pt $I_L$ to $I_S$ conversion layer thickness. The thickness of the capping layer is kept the same as that of the conversion layer. **b)** The V-layer thickness dependence of the $\xi_{DL}^{E}$. A fitting using Equation 2 is used to quantify the effective orbital Hall conductivity and diffusion length of V. The insets of **(a-b)** are the schematics of the respective layer stacks. **c)** OHC $\sigma_{zx}^{L_y}$ of bcc V plotted as a function of the Fermi energy shift, separated into its LC, IC, and LC + IC contributions, and compared to the ACA.

**Table 2.** Values of the modern and ACA OHC $\sigma_{zx}^{Ly}$ of bcc V at the Fermi energy.

| Contribution | OHC [($\hbar/2e$) ($\times 10^5\ \Omega^{-1}\mathrm{m}^{-1}$)] |
|---|---|
| ACA | 9.66 |
| LC | 1.39 |
| IC | –2.62 |
| LC + IC | –1.23 |

## 3. Conclusions

In conclusion, we demonstrate a negative OHE of intrinsic origin in our V films, accompanied by substantial effective OHC and a large orbital diffusion length. Our complementary measurement schemes spanning quasi-static to ps time scales reveal that the sign of $\xi_{DL}^{E}$ from V is negative, which is opposite to the conventional theoretical predictions. Moreover, the magnitude of $\xi_{DL}^{E}$ from V depends on the choice of the FM layer. These observations, together with consistent agreement in $\xi_{DL}^{E}$ (and peak-to-peak THz emission amplitude) across a broad frequency range, confirm a negative OHE in our V films that is of intrinsic origin. Furthermore, our V samples exhibit a high $\sigma_{OH}^{eff}$ of –(1.44±0.34) ($\hbar/2e$) ($\times 10^5\ \Omega^{-1}\mathrm{m}^{-1}$), accompanied by a large $\lambda_{OH}$ of (15.0±2.5) nm. Our novel theoretical framework that captures the contributions from local as well as itinerant circulation contributions, related to the surface of the sample, to the OHE, evidences an excellent agreement between the experimentally observed $\sigma_{OH}^{eff}$ and theoretically calculated OHC (–1.23 ($\hbar/2e$) ($\times 10^5\ \Omega^{-1}\mathrm{m}^{-1}$)). This highlights the important role of itinerant circulation contributions for a complete physical understanding of intrinsic OHE. Our findings of large, negative, and intrinsic OHC in V across a broad frequency range, together with an excellent agreement with new theoretical models, underscore the critical role of itinerant circulations in the fundamental understanding of intrinsic OHE and pave the way for energy-efficient electronic devices using light metals.

## 4. Methods

*Sample Deposition and Device Fabrication:* We deposit the samples using an industrial Singulus Rotaris sputtering tool, with a base pressure in the order of $10^{-8}$ mbar. In this study, we sputter two types of samples. For 2nd harmonic Hall voltage measurements, we utilize the samples with perpendicular magnetic anisotropy (PMA) (SI S3). However, for ST-FMR and THz emission measurements, we use samples with in-plane anisotropy (see SI S1 for sample list).

For SHH voltage measurements, we fabricate standard Hall bar devices using MicroWriter optical lithography and ion beam etching (IBE) techniques. The width and length of the devices are maintained at 5 μm and 50 μm, respectively.

For ST-FMR measurements, we prepare co-planar waveguides using a two-step fabrication process. First, we fabricate a rectangular device with a width and length of 20 μm and 45 μm, respectively, using optical lithography and IBE. Subsequently, we perform another step of lithography, deposition of electrode materials, and the lift-off processes to fabricate the ground-source-ground (G-S-G) design. The electrode materials include Cr(5 nm)/Au(200 nm).

*Electrical Transport Measurements:* For SHH voltage measurements, the samples are first wire-bonded on a specific sample holder and then loaded into a vector cryostat. Further, we apply an alternating current at 133 Hz along the longitudinal direction (i.e., along the $x$-axis, Figure 1a) of the Hall cross using a Keithley 6221 current source. Subsequently, two lock-in amplifiers (SR7225 and SR7265) are utilized to simultaneously measure the 1$^{st}$ and 2$^{nd}$ harmonic components of the transverse Hall voltage. The measured data are then fitted to calculate $\xi_{DL}^{E}$ (SI S2).

For ST-FMR measurements, the samples are placed on a probe station, equipped with RF G-S-G pico-probes and an electromagnet. The RF signal from an Anritsu 68087C signal generator is amplitude-modulated using an external RF switch (RFSPSTA5M43G). Modulation pulses at a frequency of 409 Hz are generated by an external arbitrary waveform generator (Agilent 33250A). The modulated RF current is then passed through the sample, and the resulting rectified DC voltage is measured using a lock-in amplifier (SR830). We perform the measurements for all the samples in a frequency range of 8-12 GHz. The in-plane magnetic field ($B_{\mathrm{ext}}$) is swept in a range starting from 400 mT to 0 mT. Additionally, the power is kept at 10 dBm, and the angle between $B_{\mathrm{ext}}$ and current flow was fixed at 45° (& 225°). Similarly, the measured data are then fitted using the methodology detailed in the paper and SI S5. Note, $\xi_{\mathrm{DL}}^{J}$ values presented in the paper represent averages over the full frequency range.

The resistivities of all the samples are measured using the standard four-point probe method. All the measurements in this study are performed at room temperature.

*THz Emission Measurements:* During the THz emission measurements, we excite the samples using linearly polarized laser pulses generated from a Ti:Sa regenerative amplifier system. The corresponding central wavelength, pulse duration, and repetition rate are ~800 nm, ~50 fs, and ~500 Hz, respectively. The emitted THz radiation is detected in transmission geometry through electro-optic sampling using a 1-mm-thick <110> ZnTe crystal. The measurements are performed in the presence of an external magnetic field of ~140 mT and in a dry $N_2$ environment to avoid absorption losses of the THz radiation due to water.

We perform the measurements both (*i*) by rotating the direction of $B_{\mathrm{ext}}$ and (*ii*) by exciting the samples from the front (film surface) and back (substrate) sides. From the measured THz emission spectra, we focus on the odd-in magnetization ($S_{\mathrm{odd}}$ ($t$)) signal to account for THz emission from the magnetic effects[64]. Here, front and back excitations refer to the laser pulse being incident on the film surface and substrate, respectively. We also estimate the even-in magnetization ($S_{\mathrm{even}}$ ($t$)) signal, which is less than 5% of the $S_{\mathrm{odd}}$ ($t$)[64]. Subsequently, we normalize the $S_{\mathrm{odd}}$ ($t$) signal concerning pump absorptance and sample impedance to isolate the THz emission from the spin-orbitronic effects[29,68,69] (SI S7).

**Acknowledgements**

N.V., D.K., A.D., R.G., G.J., S.K., and M.K. thank the DFG (Spin+X (A01, A11, B02) TRR 173-268565370 and Project No. 358671374; the European Union's Horizon 2020 research and innovation programme under grant agreement No 863155 (s-Nebula); the European Research Council (ERC) under the European Union's Horizon 2020 research and innovation programme (Grant No. 856538, project "3D MAGiC"); the Research Council of Norway through its Centers of Excellence funding scheme, Project No. 262633 "QuSpin"; the European Union's Horizon Europe programme (EU HORIZON-CL4-2021-DIGITAL-EMERGING-01-14 under grant agreement No. 101070290 (NIMFEIA); and the EU HORIZON-EIC-2023-PATHFINDEROPEN-01-01 under grant agreement No. 101129641 (OBELIX)). R.G. acknowledges the Swedish Research Council (VR) for the International Postdoc VR grant (Grant ID: 2023-06605). D.J. and P.M.O. acknowledge support by the Swedish Research Council (VR), the Knut and Alice Wallenberg Foundation (Grants No. 2022.0079 and 2023.0336), and the Wallenberg Initiative Materials Science for Sustainability (WISE) funded by the Knut and Allice Wallenberg Foundation.

**Author contributions**

M.K. proposed and supervised the study. D.K., R.G., and G.J. deposited the thin film samples, and N.V. and D.K. fabricated the devices with inputs from S.K.. N.V., D.K., and A.D. performed the measurements and analyzed the data with the crucial inputs from M.K., S.K., G.J., L.G., Z.X., and J.O.L.. H.I.W. provided further inputs during the THz emission measurements. M.S., D.G., and Y.M. provided the theoretical support with inputs from D.J. and P.M.O.. The manuscript was written by D.K., N.V., and A.D. with the support of M.S., S.K., Y.M., and M.K.. All the co-authors commented on the manuscript.

**Data Availability Statement**

Data will be available from the corresponding author upon a reasonable request.

**Conflict of Interest Disclosure**

The authors declare no conflict of interest.